\documentstyle[aps,prl,epsf,twocolumn,floats]{revtex}
\begin{document}
\draft
% start of wide text
\twocolumn[\hsize\textwidth\columnwidth\hsize\csname @twocolumnfalse\endcsname

\title{Specific heat of Ce$_{\bf 0.8}$La$_{\bf 0.2}$Al$_{\bf3}$ in magnetic
fields: a test of the anisotropic Kondo picture}
\author{R.\ Pietri, K.\ Ingersent, and B.\ Andraka}
\address{Department of Physics, University of Florida, \\
P.O. Box 118440, Gainesville, FL 32611-8440}
\date{July 6, 2000}
\maketitle

\begin{abstract}

The specific heat $C$ of Ce$_{0.8}$La$_{0.2}$Al$_3$ has been measured as a
function of temperature $T$ in magnetic fields up to 14\,T. A large peak
in $C$ at 2.3\,K has recently been ascribed to an anisotropic Kondo effect in
this compound. A 14-T field depresses the temperature of the peak by only
0.2\,K, but strongly reduces its height.  The corresponding peak in $C/T$
shifts from 2.1\,K at zero field to 1.7\,K at 14\,T.  The extrapolated
specific heat coefficient $\gamma=\lim_{T\rightarrow 0}C/T$ increases with
field over the range studied.  We show that these trends are inconsistent with
the anisotropic Kondo model.

\end{abstract}


\bigskip
% end of wide text
]
\narrowtext

CeAl$_3$ occupies a particularly important position in the history of heavy
fermions.  The first report of its unusual low-temperature specific
heat\cite{Andres} a quarter of a century ago led to enormous interest in this
and similar systems based on 4$f$ or 5$f$ elements.  For years, CeAl$_3$ was
considered a canonical heavy fermion system, and it greatly influenced
theoretical work in the field.  Indeed, CeAl$_3$ was viewed as a
realization of a ``standard model'' based on the Kondo effect and Fermi-liquid
theory. The hallmark properties of this compound are its specific heat and
its electrical resistivity.  The low-temperature specific heat $C(T)$ is
greatly enhanced over that for a conventional metal, with a linear coefficient
$\gamma=\lim_{T\rightarrow0}C/T\approx 1250$\,mJ/K$^2$Ce\,mol\cite{C-norm}.
The resistivity above 10\,K is described to very high very accuracy by a
theory for a Kondo impurity in crystalline electric fields\cite{Coqblin}.
Below 300\,mK the resistivity has a Fermi-liquid form $\rho = \rho_0 + AT^2$,
with a strongly enhanced $A$ coefficient.  The ratio $A/\gamma^2$
is about $10^{-5}\,\Omega$\,cm\,K$^2$mol$^2$J$^{-2}$, a value close to
that theoretically predicted for nonmagnetic Kondo lattices\cite{Moriya}, and
experimentally observed\cite{KW} in many other heavy fermion compounds.

The description of CeAl$_3$ in terms of a nonmagnetic Kondo lattice and a heavy
Fermi liquid has been challenged by microscopic measurements such as muon spin
resonance ($\mu$SR)\cite{Barth} and nuclear magnetic resonance
(NMR)\cite{Gavilano}.  According to these measurements, either short-range
magnetic order or strong antiferromagnetic correlations exist below 2\,K.  In
addition, the specific heat itself has an unexplained feature: a maximum in
$C/T$ near 0.4\,K.   A similar maximum is found in another heavy fermion
system, CeCu$_2$Si$_2$, around the same temperature.  In both compounds, this
feature was initially attributed to coherence in the Kondo lattice.  However,
extensive studies of CeCu$_2$Si$_2$ gave rise to an alternative explanation
based on weak magnetic ordering of heavy quasiparticles\cite{Bredl+Andraka}. A
previous alloying study of CeAl$_3$ has similarly pointed to a magnetic origin
for the 0.4\,K anomaly\cite{Andraka95}.  When La is partially substituted for
Ce in Ce$_{1-x}$La$_x$Al$_3$, this weak feature, observable in $C/T$ but not in
$C$ for $x=0$, gradually evolves for $x\geq 0.05$ into a large peak in both $C$
and $C/T$.  The highest-La-content alloy investigated in
Ref.~\onlinecite{Andraka95}, Ce$_{0.8}$La$_{0.2}$Al$_3$, has a pronounced
maximum in $C$ near 2.3\,K and a corresponding peak in the susceptibility at
2.5\,K, reminiscent of an antiferromagnetic transition. The smooth and
monotonic increase with $x$ of the temperature position and magnitude of the
anomaly suggests that this feature has a common origin in pure and La-doped
CeAl$_3$. The apparent enhancement of the magnetic character of the anomaly
upon La doping is consistent with Doniach's Kondo necklace model\cite{Doniach},
since doping increases the lattice constants, and therefore decreases the
hybridization between $f$ and ligand states.
However, an interpretation based on this model is somewhat undermined by
the fact that anomalies in $C$, similarly pronounced to those produced
by La substitution, can be induced by replacing Al atoms with either larger or
smaller atoms\cite{Lenkewitz}.

Recent neutron scattering and $\mu$SR studies by Goremychkin {\it et al.}
\cite{Goremychkin00} on Ce$_{0.8}$La$_{0.2}$Al$_3$ revealed the absence of
magnetic Bragg peaks, and estimated the upper limit of any possible ordered
moment to be $0.05\mu_{\mbox{\tiny B}}$.  The response function deduced from
time-of-flight measurements changes from a quasi-elastic form to an inelastic
form around 3\,K, the temperature range where features develop in the specific
heat and the magnetic susceptibility.  This result was attributed to
weakly dissipative dynamics consistent with the anisotropic Kondo model
(AKM)\cite{Costi96}. $\mu$SR spectra showed Lorentzian damping, with a
temperature-dependent damping rate that diverges also around 3\,K. The
divergence was attributed to the development of static magnetic correlations,
indicating the possibility of magnetic order of small moments, as seen in other
heavy fermion systems\cite{Broholm}.

In order to investigate further the applicability of the AKM to
Ce$_{0.8}$La$_{0.2}$Al$_3$, and to search for any contribution to the specific
heat from small-moment ordering, we have studied the effect of magnetic fields
up to 14\,T on the linear coefficient $\gamma$ and on the temperatures $T_M$
and $T_m$ of the maxima in $C$ and $C/T$, respectively.
We compare our data with numerical results for the specific heat of the AKM.
The experimental data show qualitatively different trends from the model
calculations, thereby casting considerable doubt on the validity of this
theoretical description of Ce$_{0.8}$La$_{0.2}$Al$_3$.

Our measurements used a polycrystal of Ce$_{0.8}$La$_{0.2}$Al$_3$ from a
previous alloying study\cite{Andraka95}.  The sample was prepared in an arc
furnace under an argon atmosphere, and annealed at 830$\,\mbox{}^{\circ}$C for
3 weeks. Magnetic susceptibility and x-ray diffraction measurements revealed no
sign of the secondary phases CeAl$_2$ and Ce$_3$Al$_{11}$.

\begin{figure}[t]
\centering
\vbox{\epsfxsize=80mm \epsfbox{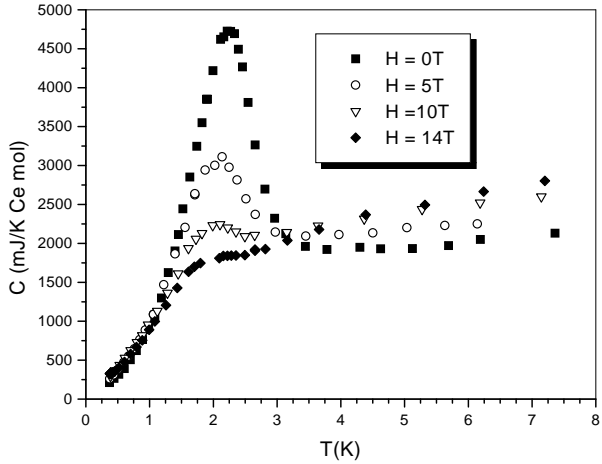}}
\caption{$C$ vs $T$ for Ce$_{0.8}$La$_{0.2}$Al$_3$ at $H=0$, 5, 10, and 14\,T.}
\label{fig1}
\end{figure}

\begin{figure}[t]
\centering
\vbox{\epsfxsize=80mm \epsfbox{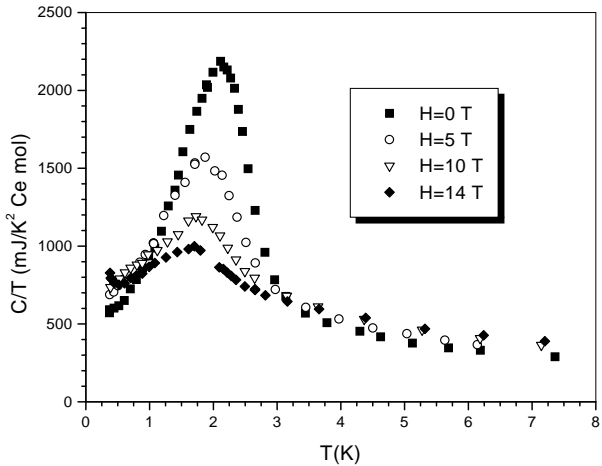}}
\vspace{1ex}
\caption{$C/T$ vs $T$ for Ce$_{0.8}$La$_{0.2}$Al$_3$ at $H=0$, 5, 10, and
14\,T.}
\label{fig2}
\end{figure}

\begin{figure}[t]
\centering
\vbox{\epsfxsize=80mm \epsfbox{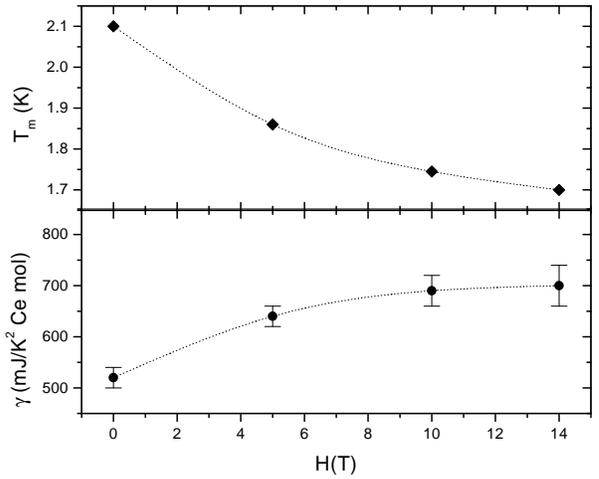}}
\vspace{1ex}
\caption{$T_m$ vs $H$ and $\gamma$ vs $H$ for
Ce$_{0.8}$La$_{0.2}$Al$_3$, where $T_m$ is the temperature of the maximum in
$C/T$.  The dotted lines are guides to the eye.}
\label{fig3}
\end{figure}

Figure~\ref{fig1} shows the specific heat of this alloy in fields of 0, 5, 10,
and 14\,T. The phonon contribution has been subtracted using the specific heat
of LaAl$_3$\cite{Edelstein}, and the remainder has been normalized to a mole of
Ce.  The same data are plotted as $C/T$ vs $T$ in Fig.~\ref{fig2}. The
main effect of the field is a strong reduction in the magnitude
of the anomalies in $C$ and $C/T$. Also striking is the very weak field
dependence of the temperature position of the anomalies. A pronounced peak
in $C$ located at $T_M \simeq2.3$\,K for $H=0$ is replaced by a shoulder
near 2.1\,K for $H=14$\,T.
The peak in $C/T$ also shifts slowly with field, $T_m$ decreasing from
2.1\,K at 0\,T to 1.7\,K at 14\,T (see Fig.~\ref{fig3}). Note that the
difference between $T_M$ and $T_m$ grows with applied field. A difference
of the same order has been observed in zero field for Ce$_{1-x}$La$_x$Al$_3$
alloys with $x< 0.2$, where $T_m\!-\!T_m$ grows as $x$ becomes
smaller\cite{Andraka95}. In this respect, an increase in the magnetic field
has a similar effect to a decrease in~$x$.

Another important result is an increase with field of $C/T$ values at low
temperatures (below 1\,K), signaling a partial restoration of the heavy
fermion state present in pure CeAl$_3$. It may be that the large nuclear
moments of Al contribute to enhance $C/T$ at the lowest temperatures and the
largest fields. Indeed, the 14-tesla $C/T$ data display a low-temperature tail
which might be due to a nuclear hyperfine contribution $\Delta C/T\propto
1/T^3$.  None of the curves at lower fields show a similar upturn.  Therefore,
the linear specific heat coefficient $\gamma$ was extracted
from a linear fit to $C/T$ vs $T^2$ below 1\,K, except for the 14-T data,
where $\gamma$ was determined from the slope of $CT^2$ vs $T^3$ below 1\,K.
As may be seen in Fig.~\ref{fig3}, $\gamma$ seems to saturate in the range
$H\gtrsim 10$\,T.  (The error bars for $\gamma$ combine experimental
and regression uncertainties.)

\begin{figure}[t]
\centering
\vbox{\epsfxsize=80mm \epsfbox{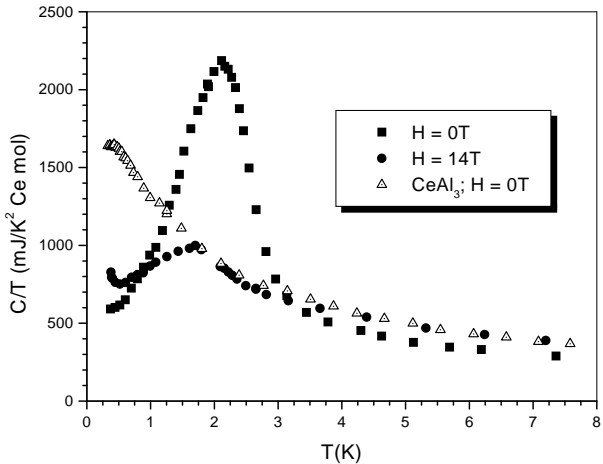}}
\vspace{1ex}
\caption{$C/T$ vs $T$ for Ce$_{0.8}$La$_{0.2}$Al$_3$ at $H=0$ and 14\,T, and
for CeAl$_3$ at $H=0$.}
\label{fig4}
\end{figure}

It is worth noting that $C/T$ for Ce$_{0.8}$La$_{0.2}$Al$_3$ at 14\,T and $C/T$
for CeAl$_3$ in zero field coincide above 4\,K to within the accuracy of the
measurement.  This is demonstrated in Fig.~\ref{fig4}, which also includes the
corresponding curve for Ce$_{0.8}$La$_{0.2}$Al$_3$ at $H=0$.  Since $C/T$ for
the pure compound is only weakly field dependent above 4\,K (for fields
$\sim 10$\,T)\cite{unp}, we can claim that the high-field ($H\sim 14$\,T)
specific heats for these two alloys converge in this temperature regime.

We now attempt to analyze our magnetic-field data in terms of the anisotropic
Kondo model (AKM) for a single magnetic impurity. The model assumes an exchange
interaction $J_z S_z s_z + J_{\perp} (S_x s_x + S_y s_y)$ between the impurity
spin~$\bf S$ and the net conduction-electron spin~$\bf s$ at the impurity site.
Goremychkin {\it et al.}\cite{Goremychkin00} have proposed the AKM as a
description for the thermodynamic properties of both Ce$_{0.8}$La$_{0.2}$Al$_3$
and CeAl$_3$.  A strong dependence on field orientation in the magnetic
susceptibility of CeAl$_3$ single crystals\cite{Jaccard} is suggestive
of anisotropic behavior corresponding to $J_z\!\gg\!J_{\perp}\!>\!0$, with the
magnetic $z$~direction being the crystallographic $c$~axis.

The AKM is known to be equivalent in the limit of low energies to a number of
other models. In recent years a mapping\cite{Chakravarty82}
onto the spin-boson model with Ohmic dissipation has been exploited
\cite{Costi96,Costi98+99} to deduce physical properties of dissipative
two-level systems\cite{Leggett87} from numerical calculations performed for the
AKM. These studies have shown that under certain conditions, the impurity
contribution to the zero-field heat capacity of the AKM exhibits peaks both in
$C$ and $C/T$, qualitatively similar to the data plotted in Figs.~\ref{fig1}
and~\ref{fig2}. Such peaks are found only for $\varrho_0 J_z\gtrsim
1$, where $\varrho_0$ is the conduction-band density of states at the Fermi
energy. For a given value of $\varrho_0 J_z$, the specific heat for
all $J_{\perp} \ll J_z$ can be collapsed onto a universal scaling
curve $C/\gamma T$ vs $T/T_m$\cite{Costi98+99}. Moreover, the temperature of
the peak in $C/T$ is given by $T_m=\alpha R / \gamma$ \cite{C-norm}, where $R$
is the gas constant and $\alpha$ is a function of $\varrho_0 J_z$
(only).

\begin{figure}[t]
\centering
\vbox{\epsfxsize=80mm \epsfbox{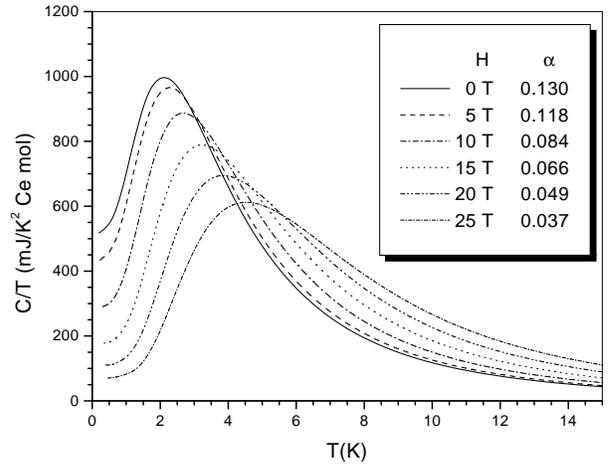}}
\vspace{1ex}
\caption{$C/T$ vs $T$ calculated for the anisotropic Kondo model in various
magnetic fields $H$, with model parameters chosen so that $\alpha=0.130$ for
$H=0$. See text for details.}
\label{fig5}
\end{figure}

From the observed peak position $T_m=2.1$\,K and a value
$\gamma=520$\,J/K$^2$Ce\,mol extracted as described above, we deduce
$\alpha=\gamma T_m/R= 0.13$ for Ce$_{0.8}$La$_{0.2}$Al$_3$ in zero field,
in agreement with the estimate of Ref.~\onlinecite{Goremychkin00}.
We have used $\alpha$ and $T_m$ as inputs for a numerical renormalization-group
calculation\cite{NRG} of the specific heat of the AKM. Figure~\ref{fig5}
shows the predicted effect of applying a uniform magnetic field along the
magnetic $z$~axis, under the assumption that the impurity and the conduction
electrons have $g$ factors $g_i = g_e = 2$. (Changing the $g$ factors
multiplies the field scale by an overall factor, but does not otherwise affect
the results\cite{field_scale}.)

The numerical data exhibit three main trends with increasing field:
(1)~The anomaly in $C/T$ becomes broader and lower.
(2)~The peak shifts markedly to higher temperatures.
(3)~$C/T$ decreases at all temperatures below the zero-field value of $T_m$;
the fractional change in $\gamma$ is greater than that in the peak height,
so that $\alpha = \gamma T_m / R$ decreases monotonically with increasing
magnetic field, as shown in the legend of Fig.~\ref{fig5}.

These numerical results are directly applicable only to single-crystal
Ce$_{0.8}$La$_{0.2}$Al$_3$ with a magnetic field along the $c$~axis. For
comparison with our polycrystalline data, one must average over all possible
field orientations. The Ising-like crystal-field ground state of Ce$^{3+}$ in
CeAl$_3$ \cite{Goremychkin99} implies that $g_i\!=\!0$ for the basal-plane
components of the magnetic field and, hence, that the specific heat of a
polycrystal in field $H$ is an equally weighted average of the single-crystal
results for all fields between zero and $H$. This averaging process
preserves trends (1)--(3) above.

Trend (1) accords well with our measurements, but (2) and (3) both run
counter to experiment. In Ce$_{0.8}$La$_{0.2}$Al$_3$, $T_m$ does not rise
with increasing field, but instead is weakly depressed, while $C/T$ undergoes
a small increase at temperatures much below $T_m$.  In particular, $\gamma$
rises sufficiently fast that $\alpha$ remains essentially constant up to a
14\,T field (see Table~I), in contrast to the prediction of the AKM.

The preceding comparisons seem to indicate significant shortcomings in the AKM
as a description of Ce$_{0.8}$La$_{0.2}$Al$_3$ in magnetic fields. One reason
for the inadequacy of the AKM may be the neglect of magnetic correlations
around the temperature of the maximum, as identified in the $\mu$SR studies of
Ref.~\onlinecite{Goremychkin00}.  It was noted above that the
specific heat anomaly is reminiscent of an antiferromagnetic phase transition.
The entropy under the peak in $C/T$ is a large fraction ($\sim$50$\%$) of $R\ln
2$, and the linear specific heat coefficient $\gamma\!=\!520$\,mJ/K$^2$Ce\,mol
is less than half that of pure CeAl$_3$ ($\gamma\!=\!1250$\,mJ/K$^2$Ce\,mol).
However, our field data suggest that any magnetic transition associated with
the anomaly is rather unusual.  We find that $T_M$ and $T_m$ are depressed
in an applied field at a much lower rate than is the N\'{e}el temperature in
Ce-based heavy fermion systems that order antiferromagnetically.
In CeCu$_{5.2}$Ag$_{0.8}$, for example, $T_N$ is reduced from 0.7\,K
to 0\,K in a field of about 2.5\,T\cite{Heuser}. In CePb$_3$, which exhibits
unconventional small-moment ordering at 1.1\,K, a field of order 10\,T
depresses $T_N$ to zero\cite{Julian}.

\begin{table}
\caption{Values of the specific heat coefficient $\gamma$, the peak
temperature $T_m$, and $\alpha = \gamma T_m/R$ (where $R$ is the gas constant)
for Ce$_{0.8}$La$_{0.2}$Al$_3$ in different magnetic fields $H$.}
\vspace{1ex}
\label{table1}
\begin{tabular}{cccc}
$H$(T)& $\gamma$ (mJ/K$^2$ Ce\,mol)& $T_m$(K)& $\alpha$\\
\tableline
\tableline
0& $520\pm20$ & $2.13\pm0.02$ & $0.133\pm0.005$ \\
5& $640\pm20$ & $1.86\pm0.02$ & $0.143\pm0.005$ \\
10& $690\pm30$ & $1.75\pm0.02$ & $0.145\pm0.006$ \\
14& $700\pm40$ & $1.70\pm0.02$ & $0.143\pm0.008$ \\
\end{tabular}
\end{table}

In summary, we have measured the heat capacity of Ce$_{0.8}$La$_{0.2}$Al$_3$ as
a function of temperature in magnetic fields up to 14\,T. The field strongly
diminishes the peaks found around 2\,K in both $C$ and $C/T$, but only weakly
depresses the peak temperatures. The linear specific heat coefficient
increases with field in the direction of the value for pure CeAl$_3$,
implying partial restoration of the heavy fermion state suppressed
by La doping.  We have analyzed our data in terms of the anisotropic Kondo
model.  The model predicts a shift of the peak in $C/T$ to higher temperatures
with increasing field, accompanied by a significant reduction in $C/T$
at low temperatures.  These two trends are at odds with experiment.  Our
results do not rule out an alternative theoretical picture based on
small-moment magnetism.  However, the field-insensitivity of the temperature
of the heat-capacity peak remains to be understood within this scenario.

This work was supported in part by Department of Energy grant
DE-FG02-99ER45748.


\begin{thebibliography}{29}

\bibitem{Andres}
K.\ Andres, J.\ E.\ Graebner, and H.\ R.\ Ott, Phys.\ Rev.\ Lett.\ {\bf 27},
1779 (1975).

\bibitem{C-norm}
Throughout this paper, $C$ and $\gamma$ are assumed to be normalized per mole
of Ce.

\bibitem{Coqblin}
B.\ Cornut and B.\ Coqblin, Phys.\ Rev.\ B {\bf 5}, 4541 (1972).

\bibitem{Moriya}
T.\ Takimoto and T.\ Moriya, Solid State Commun.\ {\bf 99}, 457 (1996).

\bibitem{KW}
K.\ Kadowaki and S.\ B.\ Woods, Solid State Commun.\ {\bf 58}, 507 (1986).

\bibitem{Barth}
S.\ Barth, H.\ R.\ Ott, F.\ N.\ Gygax, B.\ Hitti, E.\ Lippelt,
A.\ Schenck, C.\ Baines, B.\ van den Brandt, T.\ Konter, and S.\ Mango,
Phys.\ Rev.\ Lett.\ {\bf 59}, 2991 (1987).

\bibitem{Gavilano}
J.\ L.\ Gavilano, J.\ Hunziker, and H.\ R.\ Ott, Phys.\ Rev.\ B {\bf 52},
13106 (1995).

\bibitem{Bredl+Andraka}
C.\ D.\ Bredl, S.\ Horn, F.\ Steglich, B.\ L\"{u}thi, and R.\ M.\ Martin,
Phys.\ Rev.\ Lett.\ {\bf 52}, 1982 (1984);
B.\ Andraka, G.\ R.\ Stewart, and F.\ Steglich, Phys.\ Rev.\ B {\bf 48},
3939 (1993).

\bibitem{Andraka95}
B.\ Andraka, C.\ S.\ Jee, and G.\ R.\ Stewart, Phys.\ Rev.\ B {\bf 52},
9462 (1995); R.\ Pietri and B.\ Andraka, Physica B {\bf 230-232}, 535 (1997).

\bibitem{Doniach}
S.\ Doniach, in {\it Valence instabilities and related narrow band
phenomena}, ed.\ by R.\ D.\ Parks (Plenum Press New York), p.\ 169 (1977);
S.\ Doniach, Physica B {\bf 91}, 231 (1977).

\bibitem{Lenkewitz}
S.\ Corsepius, M.\ Lenkewitz, and G.\ R.\ Stewart, J.\ Alloys Comp.\
{\bf 259}, 29 (1997).

\bibitem{Goremychkin00}
E.\ A.\ Goremychkin, R.\ Osborn, B.\ D.\ Rainford, and A.\ P.\ Murani, Phys.\
Rev.\ Lett.\ {\bf 84}, 2211 (2000).

\bibitem{Costi96}
T.\ A.\ Costi and C.\ Kieffer, Phys.\ Rev.\ Lett.\ {\bf 76}, 1683 (1996).

\bibitem{Broholm}
C.\ Broholm, J.\ K.\ Kjems, W.\ J.\ L.\ Buyers, P.\ Matthews,
T.\ T.\ M.\ Palstra, A.\ A.\ Menovsky, and J.\ A.\ Mydosh,
Phys.\ Rev.\ Lett.\ {\bf 58} 1467 (1987).

\bibitem{Edelstein}
A.\ S.\ Edelstein, R.\ A.\ Fischer, and N.\ E.\ Phillips, J.\ Appl.\ Phys.\
{\bf 61}, 3177 (1987).

\bibitem{unp}
B.\ Andraka and G.\ R.\ Stewart (unpublished).

\bibitem{Jaccard}
D.\ Jaccard, R.\ Cibin, A.\ Bezinge, J.\ Sierro, K.\ Matho,
and J.\ Flouquet, J.\ Mag.\ Mag.\ Mater.\ {\bf 76-77}, 255 (1988).

\bibitem{Chakravarty82}
S.\ Chakravarty, Phys.\ Rev.\ Lett.\ {\bf 49}, 681 (1982);
A.\ J.\ Bray and M.\ A.\ Moore, {\it ibid.} {\bf 49}, 1545 (1982);
F.\ Guinea, V.\ Hakim, and A.\ Muramatsu, Phys.\ Rev.\ B {\bf 32}, 4410 (1985).

\bibitem{Costi98+99}
T.\ A.\ Costi, Phys.\ Rev.\ Lett.\ {\bf 80}, 1038 (1998);
T.\ A.\ Costi and G.\ Zar\'{a}nd, Phys.\ Rev.\ B {\bf 59}, 12398 (1999).

\bibitem{Leggett87}
A.\ J.\  Leggett, S.\ Chakravarty, A.\ T.\ Dorsey, M.\ P.\ A.\ Fisher,
A.\ Garg, and W.\ Zwerger, Rev.\ Mod.\ Phys.\ {\bf 59}, 1 (1987);
{\bf 67}, 725(E) (1995).

\bibitem{NRG}
$C(T)$ was computed as described in C.\ Gonzalez-Buxton and K.\ Ingersent,
Phys.\ Rev.\ B {\bf 57}, 14254 (1998). We chose $\rho_0 J_{\perp}
= 10^{-3}$, and found $\rho_0 J_z = 1.90$ from the condition
$\alpha(H=0)=0.130$.  The values of $T_m$ and $\gamma=\alpha R/T_m$ fixed the
horizontal and vertical scales of Fig.~\ref{fig5}.

\bibitem{field_scale}
A magnetic field $H$ enters the partition function of the AKM in the
combination $[1-(2g_e/\pi g_i) \tan^{-1}(\pi\rho_0 J_z/4)]$ $g_i \mu_B H$;
see P.\ B.\ Vigmann and A.\ M.\ Finkel'stein, Zh.\ Eksp.\ Teor.\ Fiz.\
{\bf 75}, 204 (1978) [Sov.\ Phys.\ JETP {\bf 48}, 102 (1978)].
For fixed $\rho_0 J_z$, any change in the $g$ factors can be
treated as an effective rescaling of $H$.
In fact, the curves in Fig.~\ref{fig5} were calculated for $g_i = 2$ and
$g_e = 0$; then the field scale was converted to that for $g_i = g_e = 2$.

\bibitem{Goremychkin99}
E.\ A.\ Goremychkin, R.\ Osborn, and I.\ L.\ Sashkin, J.\ Appl.\ Phys.\
{\bf 85}, 6046 (1999).

\bibitem{Heuser}
K.\ Heuser, J.\ S.\ Kim, E.-W.\ Scheidt, T.\ Schreiner, and G.\ R.\ Stewart,
Physica B {\bf 259-261}, 392, (1999).

\bibitem{Julian}
J.\ McDonough and S.\ R.\ Julian, Phys.\ Rev.\ B {\bf 53}, 14411 (1996), and
references therein; T.\ Ebihara, K.\ Koizumi, S.\ Uji, C.\ Terakura, T.\
Terashima, H.\ Suzuki,  H.\ Kitazawa, and G.\ Kido, Phys.\ Rev.\ B {\bf 61},
2513 (2000).

\end{thebibliography}
\end{document}